\documentclass[twocolumn,prb,showpacs,multicol,amsmath,amssymb]{revtex4}
\usepackage[dvips]{graphicx}
\usepackage{graphicx}
\usepackage{dcolumn}
\usepackage{bm}
\usepackage{graphics}
\usepackage{epsfig,color}

\newcommand{\be}{\begin{equation}}
\newcommand{\ee}{\end{equation}}
\newcommand{\bea}{\begin{eqnarray}}
\newcommand{\eea}{\end{eqnarray}}

\begin{document}

\title{Quantum corrections of the biquadratic interaction in the 1D spin-1/2 frustrated ferromagnetic systems}

\author{Javad Vahedi$^1$, Saeed Mahdavifar$^2$}
\affiliation{ $^1$  Department of Physics, Science and Research
Branch, Islamic Azad University, Tehran, Iran\\
$^2$ Department of Physics, University of
Guilan,41335-1914, Rasht, Iran}
\date{\today}

\begin{abstract}
 Quantum corrections of the biquadratic interaction in the 1D spin-1/2 frustrated ferromagnetic
Heisenberg model are studied. The biquadratic interaction for spin-1/2 chains is eliminated and
transformed to the quadratic interaction. Doing a numerical experiment, new insight as to how the classical
phases get modified on the inclusion of quantum fluctuations is provided. Observed results suggest
the existence of an intermediate region in the ground state phase diagram of the frustrated
ferromagnetic spin-1/2 chains with combination of dimer and chiral orders. In addition, from
the quantum entanglement view point, differences  between quantum phases are also obtained.
The nearest neighbor spins never be entangled in the frustrated ferromagnetic chains but are
entangled up to the Majumdar-Ghosh point in the frustrated antiferromagnetic chains. On the
other hand, the next nearest neighbor spins in the mentioned intermediate region are entangled.
\end{abstract}


\maketitle


\section{Introduction}\label{sec1}
The explore of novel order in frustrated models in low dimensional
quantum systems have been studied extensively from theoretical
and experimental point of view. An example
which shows a variety of intriguing phenomena is
 frustrated ferromagnetic spin-$\frac{1}{2}$ chain with added nearest-neighbor biquadratic interaction\cite{kaplan}:
\begin{equation}
\emph{H}=\sum_{n=1}^{N}\big[J_{1}\vec{S}_{n}.\vec{S}_{n+1}+J_{2}\vec{S}_{n}.\vec{S}_{n+2}-A(\vec{S}_{n}.\vec{S}_{n+1})^{2}\big],
\label{e1}
\end{equation}
where $J_{1}<0$, $J_{2}>0$ are the nearest-neighbor (NN) and next-nearest-neighbor
(NNN) exchange couplings. $\vec{S}_{n}$ represents the spin-$\frac{1}{2}$ operator at the $n$th
site,  and $A$ denotes the biquadratic exchange. We introduce parameters $\alpha=\frac{J_{2}}{|J_{1}|}$ and $a=\frac{A}{|J_{1}|}$
for convenience.

The pure frustrated ferromagnetic model ($a=0$) is well studied\cite{ Aligia,Dmitri,Hikihara,Mahdavifar}.
Beside a general interest in understanding {\em frustrations} and phase transitions,
it helps people to understand intriguing magnetic properties of a novel class of
edge-sharing copper oxides, described by the frustrated ferromagnetic model\cite{Mizuno, Hase, Solodovnikov}.
Several compounds with edge-sharing chains are known, such as $Li_{2}CuO_{2}$,
 $La_{6}Ca_{8}Cu_{21}O_{41}$, and $Ca_{2}Y_{2}Cu_{5}O_{10}$\cite{Mizuno}.
 Though the pure frustrated ferromagnetic model has been a subject of many studies \cite{Tonegawa,
Chubukov, Cabra, Krivnov} the complete picture of the quantum phases
of this model has remained unclear up to now. It is known that the
ground state is ferromagnetic for $\alpha=\frac{J_{2}}{|J_{1}|}<\frac{1}{4}$. At $\alpha_{c}=1/4$
the ferromagnetic state is degenerate with a singlet state. The
wave function of this singlet state is exactly known \cite{Hamada, Dmitriev}.
For $\alpha>\frac{1}{4}$ however, the ground state is an incommensurate singlet. It has been long
believed that at $\alpha>\frac{1}{4}$ the model is
gapless\cite{White, Allen} but the one-loop renormalization
group analysis indicates\cite{Cabra, Nersesyan} that the gap is
open due to a Lorentz symmetry breaking perturbation. However,
existence of the energy gap has not been yet verified
numerically\cite{Cabra}. Using field theory
considerations it has been proposed\cite{Dmitriev} that a very
tiny but finite gap exists which can be hardly observed by numerical techniques.

\begin{figure}[t]
\includegraphics[width=0.95\columnwidth]{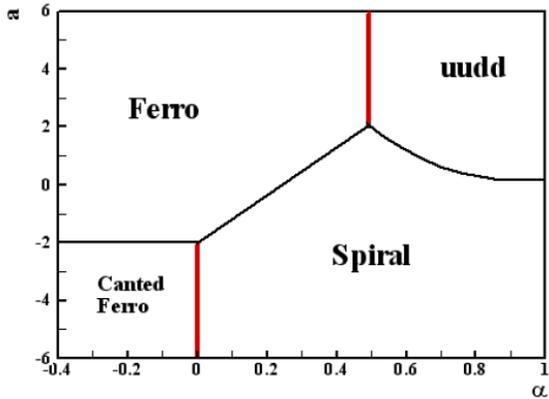}
\caption{(Color online) Classical phase diagram: $\textbf{a}\equiv A/ |J1|$ vs $\alpha\equiv J2 / |J1|$.
Disorder occurs on the emphasized vertical line segments.}
\label{schematic1}
\end{figure}

In a very recent work \cite{kaplan}, T. Kaplan presents the classical ground state phase
diagram of the frustrated model with added biquadratic  exchange interaction ($a\neq 0$).
By considering spins as vector and using a kind of cluster method which is based
on a block of three spins, he found the classical ground state phase diagram as Fig.~{\ref{schematic1}}.
The classical phase diagram exhibits the ferromagnetic, the spiral, the canted-ferro, and up-up-down-down spin structures.
In the non frustrated Heisenberg case ($\alpha=0$), the spiral phase is caused
by the contest between the Heisenberg and the biquadratic
interactions\cite{kaplan}.
There are two known sources of these terms: Firstly, purely
electronic: higher order terms in the hopping amplitudes
or orbital overlap (leading order yields the Heisenberg
interactions)\cite{bb1,bb2} and Secondly, lattice induced: spin-lattice
interaction \cite{bb3}.


The presence of chiral phase in quasi-one dimensional frustrated magnets
has been intensively studied during the last decade\cite{Nersesyan,bb11,bb12,bb13,bb14,bb15}.
This interest was triggered by the prediction
of a ground state with non-zero vector spin chirality,
$\left< \vec{S}_{l}\times\vec{S}_{m}\right>\neq0$.
As it is pointed in ref.[30], classical states with spontaneously
 broken chirality only exist together with helical long 	
range order. The helical order breaks the continuous symmetry of
global spin rotations along the $z$-axis. Consequently, 	
the existence of long range helical order is in most cases precluded by
zero point fluctuations of 1D quantum systems \cite{bb16} 	
(Mermin-Wagner theorem \cite{bb17}). On the other hand, chiral
orderings are allowed because they only break discrete symmetries.
For this reason, chiral orders in quantum spin 	
systems can be thought as remnants of the helical order in 	
classical systems. This is one of the main motivations 	
for finding chiral orders in quantum spin Hamiltonian whose 	
ground state exhibits helical order in the $S\rightarrow\infty$ limit\cite{bb16}. 		


The structure of the paper is as follows: In Sec. II we check the validity of
classical phase diagram capture exhaustively with the accurate lanczos
scheme from quantum point of view. In Sec. III we will use the entanglement
of formation (EoF) to check  the presence of quantum phase transitions and check the
presence of critical lines which were predicted by T.Kaplan approach. Finally,
we will present our results.

\section{Quantum phase diagram} \label{sec2}
By considering operator $2(S_n.S_{n+1})+1/2$ as the permutation operator, the 1D frustrated
ferromagnetic Hamiltonian is transformed to the following model
\begin{equation}
\emph{H}^{T}=\sum_{n=1}^{N}\big[-(1+\frac{a}{2})\vec{S}_{n}.\vec{S}_{n+1}+\alpha \vec{S}_{n}.\vec{S}_{n+2}\big]+constant.
\label{et}
\end{equation}
This is nothing but the isotropic spin-1/2 Heisenberg model with NN exchange $(1+\frac{a}{2})$
and NNN exchange $\alpha$. From quantum point of view, one encounter with four different cases by
changing the strength of the biquadratic and the frustration exchanges
\begin{eqnarray}
(I)&\alpha&<0,~~a<-2,~~~ nonfrustrated~AF-F~ model \nonumber\\
(II)&\alpha&<0,~~a>-2,~~~nonfrustrated~F-F~model \nonumber \\
(III)&\alpha&>0,~~a<-2,~~~frustrated~AF-AF~model \nonumber \\
(IV)&\alpha&>0,~~a>-2,~~~frustrated~F-AF~model.\nonumber
\end{eqnarray}
It is known that the ground state of the 1D spin-1/2 non-frustrated F-F model has the ferromagnetic
long-range order. On the other hand the spectrum of the non-frustrated AF-F model is gapless.
The 1D frustrated AF-AF is well known. In the classical limit the system develops spiral order
for $\frac{\alpha}{\mid1+a/2\mid}>\frac{1}{4}$ whereas a quantum phase transition into a
dimerized phase occurs at $\alpha_{c}\simeq 0.2411~\mid1+a/2\mid$. This dimerized phase is
characterized by a singlet ground state with twofold degeneracy and an excitation gap to the
first excited state. At the Majumdar-Ghosh point\cite{saeed0}, i.e. $\alpha=0.5~\mid1+a/2\mid$
the  ground state is exactly solvable. In addition, the ground state of the frustrated F-F model
is ferromagnetic for $\frac{\alpha}{1+a/2}<\frac{1}{4}$. At $\alpha_{c}=\frac{1}{4}(1+a/2)$
the ferromagnetic state is degenerate with a singlet state. For $\alpha>\alpha_{c}$, the existence
of a tiny gapped region suggested. Recently, the possible relevance of this model to the several
quasi-1D edge-sharing cuprates\cite{saeed1, saeed2, saeed3, saeed4, saeed5} is raised very
serious\cite{saeed6, saeed7, saeed8, saeed9}. These compounds can exhibit multiferroic behavior in
low-temperature chiral spin ordered phases.  Theoretically, the study of the anisotropy effect
clearly has shown that the chiral phase appears and extends up to the vicinity of the SU(2) point
 for moderate values of frustration\cite{saeed7, saeed9} in well agreement with the experimental results.

In the following, to find the  ground state quantum phase diagram and providing  proper insight as how
the classical phases can modify by the inclusion of quantum fluctuations,
we did a  numerical experiment by using the Lanczos method. To explore the nature of the spectrum and the
quantum phase transitions, we diagonalized numerically chains with length up to $N=24$ for different values
of the biquadratic exchanges. The energies of the few lowest eigenstates were
obtained for chains with periodic boundary conditions.

\begin{figure}[t]
\includegraphics[width=1.0\columnwidth]{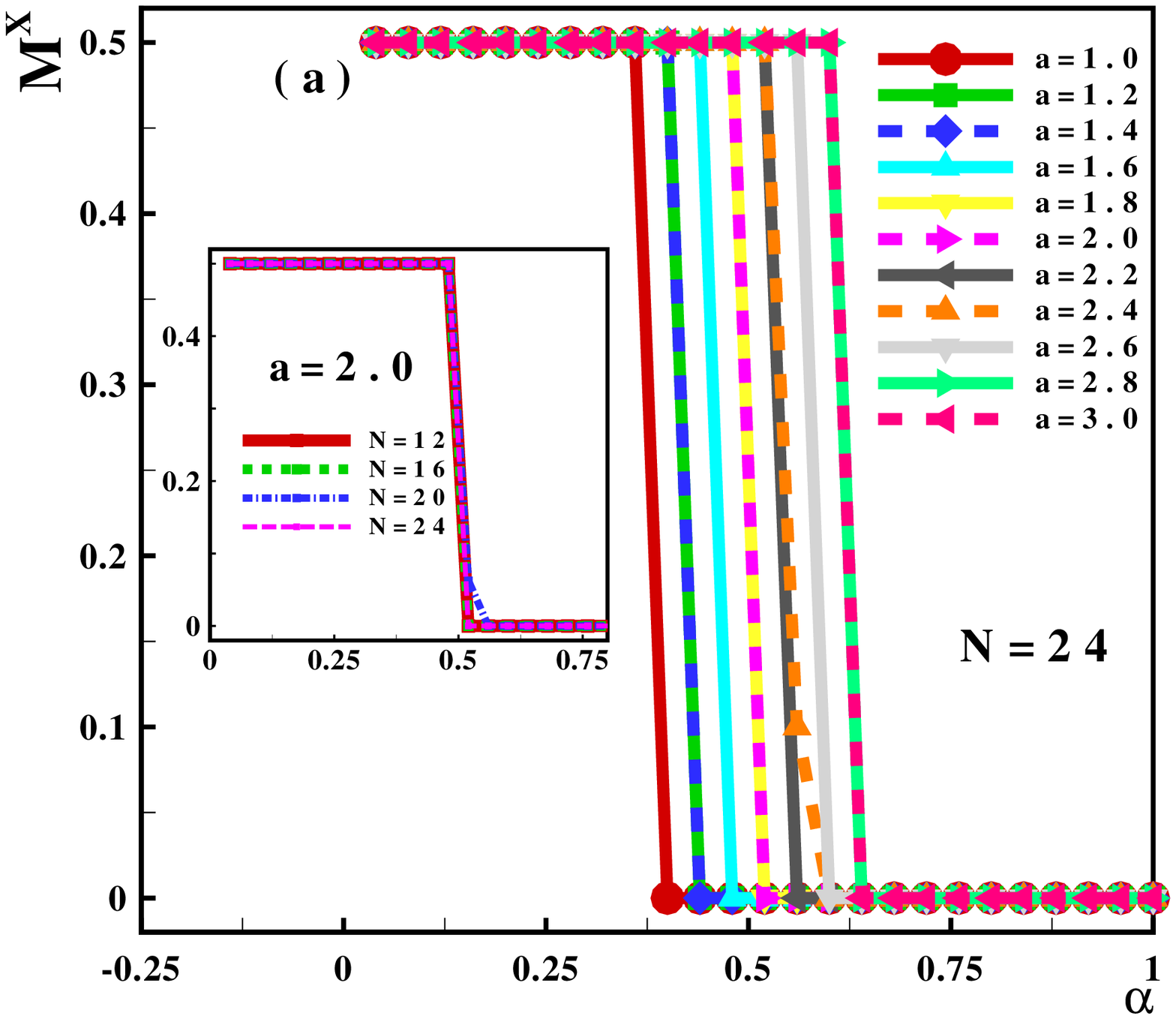}
\includegraphics[width=1.0\columnwidth]{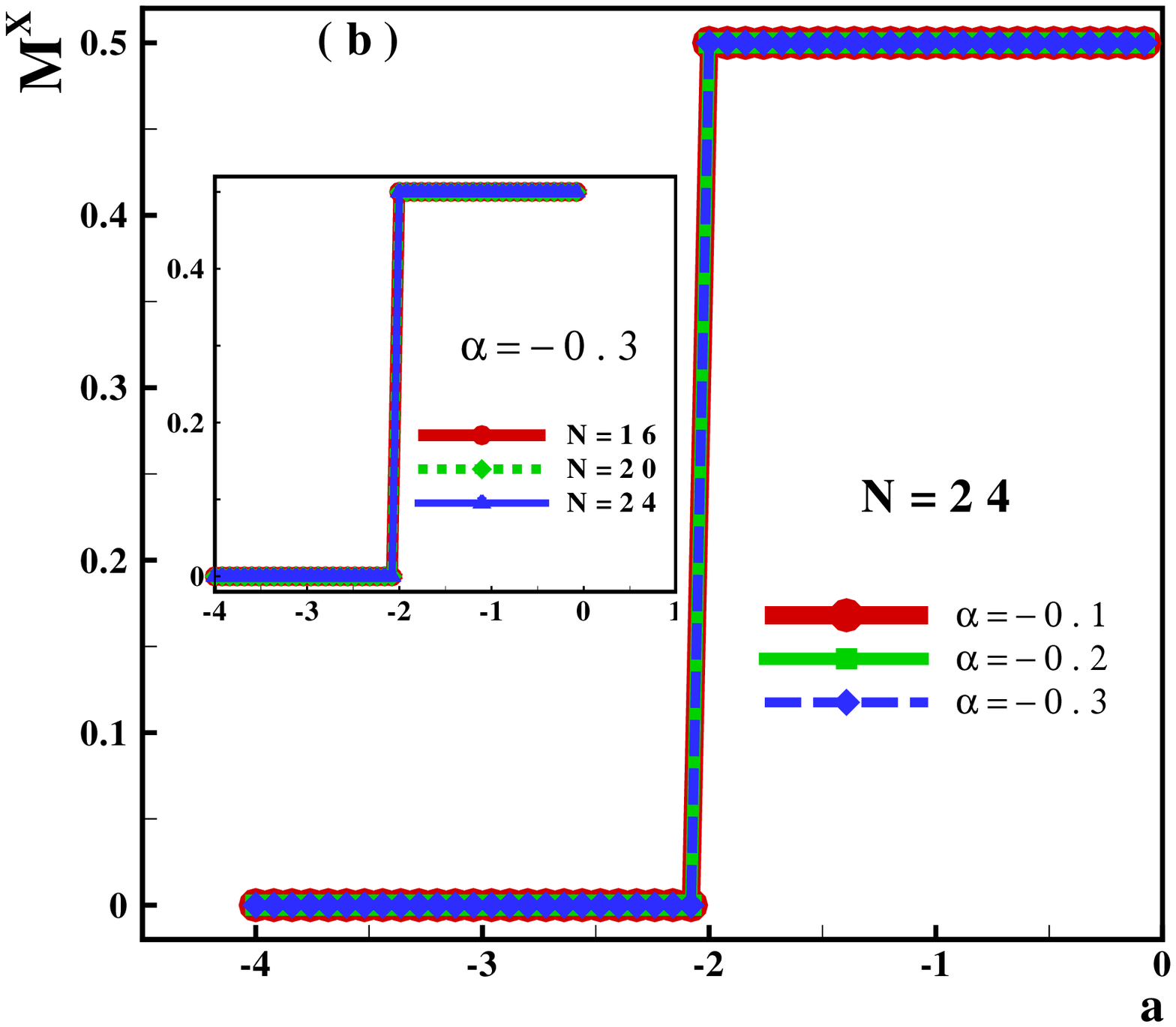}
\caption{(Color online.) Magnetization ($M^{x}$) curve versus (a)
frustration parameter $\alpha$ with different fixed biquadratic parameters \textbf{a} = 1.0, 1.2, ..., 3.0
for chain with length N =24.  (b) biquadratic parameter \textbf{a} with different fixed frustration $\alpha=-0.1, -0.2, -0.3$
for chain with length N =24. In both plots the inset shows scaling behavior for chain with lengths N= 12, 16, 20, 24.}
\label{magnetization}
\end{figure}

We start our study with magnetization where defined as
\begin{equation}
M^{\gamma}=\frac{1}{N} \sum_{j=1}^{N}\left<GS\mid S^{\gamma}_{j}\mid GS\right>
\label{e3}
\end{equation}
where $\gamma=x, y, z$ and the notation $\left<GS\mid ... \mid GS\right>$ represents
the ground state expectation value. One of the most intriguing properties of quasi-one dimensional
frustrated systems is the dependence of the magnetization on the applied magnetic
field at $T=0$. The magnetization is characterized by a
swift increase (or even discontinuity) in the magnetization when
the external field exceeds a critical value. It is expected that
the magnetization exhibits a true jump (the metamagnetic
transition) when the frustration $\alpha$ is a little larger
than $\alpha_{c}=0.25$\cite{ Aligia,Dmitriev}. In  Fig.~\ref{magnetization}(a) and Fig.~\ref{magnetization}(b),
for chain size $N=24$, we have plotted $M^{x}$  as a function of frustration and biquadratic parameters
respectively in order to sweep all parts of the ground state phase diagram. As it can be seen from
Fig.~\ref{magnetization}(a), the magnetization is saturated, $M^{x}=0.5$, in the ground state of
the nonfrustrated F-F model and for some  values of the frustration, $\alpha<\alpha_{c}=\frac{1}{4}(1+a/2)$
in the frustrated F-AF model. At the critical point $\alpha_c=\frac{1}{4}(1+a/2)$, a sudden jump is
happened which is known as the metamagnetic phase transition\cite{Mahdavifar}. Numerical results presented in
Fig.~\ref{magnetization}(b) show that quantum fluctuations destroy
  the suggested classical long range canted ferromagnetic order in the nonfrustrated AF-F model.
  By changing the biquadratic exchange a metamagnetic phase transition between nonfrustrated AF-F
  and F-F models happens at the exact critical biquadratic exchange $a=-2.0$.  In the insets of
  Fig.~\ref{magnetization} we have plotted the magnetization for a fixed value of the biquadratic
  interaction Fig.~\ref{magnetization}(a) and frustration parameter Fig.~\ref{magnetization}(b)
  for different chain sizes $N=12, 16, 20, 24$. It is completely clear that there is not any size effect on the numerical
results of the magnetization that confirms the presence of  critical lines in the thermodynamic
limit. In conclusion the quantum critical line which separates the ferromagnetic
phase from the spiral phase is consistent with the classical line, but our calculations show that the vertical critical line
which separates the ferromagnetic phase from up-up-down-down phase no longer exists in the quantum level and
quantum correlations expand the ferromagnetic phase to live even in the region $\alpha \geq 0.5$ and $a \geq 2.0$.

 \begin{figure}[t]
\includegraphics[width=1.0\columnwidth]{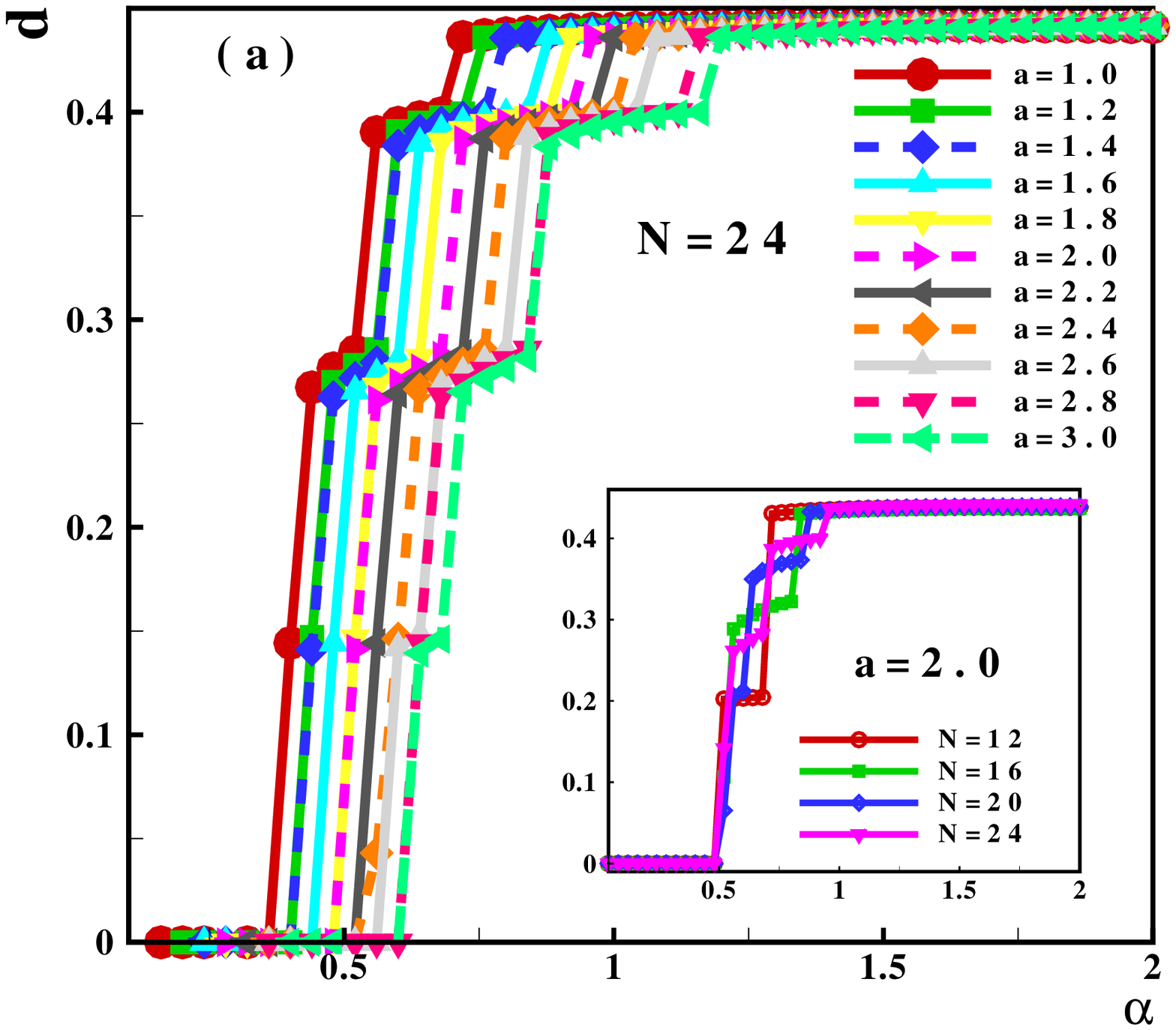}
\includegraphics[width=1.0\columnwidth]{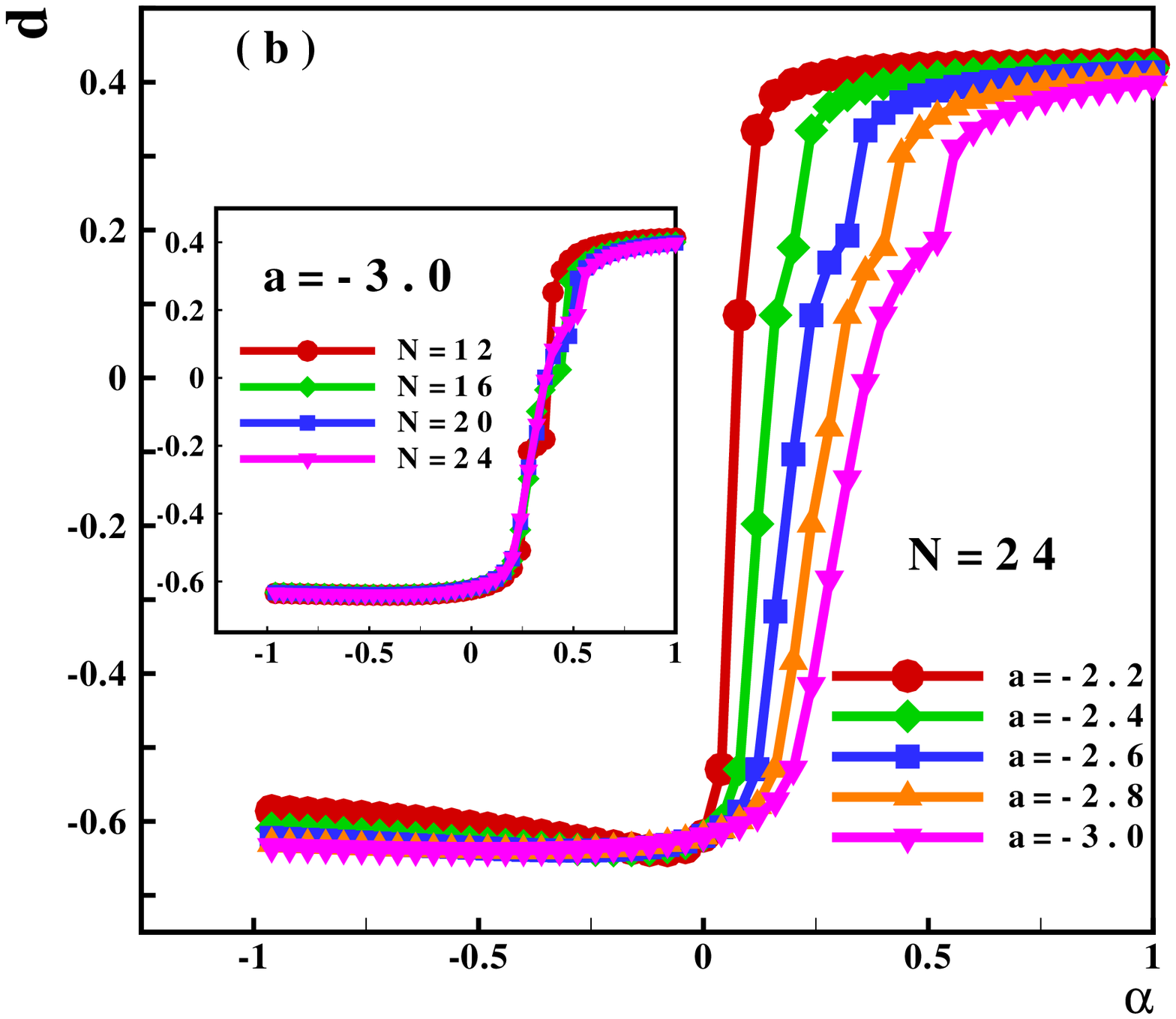}
\caption{(Color online.) The  dimer order parameter $d$ as
function of (a) frustration parameter $\alpha$ with different fixed biquadratic parameters \textbf{a} = 1.0, 1.2, ..., 3.0
for chain with length N =24.  (b) biquadratic parameter \textbf{a} with different fixed frustration $\alpha=-0.1, -0.2, -0.3$
for chain with length N =24. In both plots the inset shows scaling behavior for chain with lengths N= 12, 16, 20, 24.}
\label{dimer}
\end{figure}


To display the quantum ground state magnetic phase diagram of
the model and check the nature of the classical suggested up-up-down-down phase we
have calculated the quantum dimer order parameter which is defined as
\begin{eqnarray}
d=\frac{1}{N} \sum_{j} \langle GS|  \vec{S}_{j}\cdot\vec{S}_{j+1}-\vec{S}_{j}\cdot\vec{S}_{j+2} |GS\rangle.
\label{e4}
\end{eqnarray}
In Fig.~\ref{dimer}(a), we have plotted the dimer order parameter $d$ as a function of the frustration parameter
$\alpha$ with different fixed values of the biquadratic parameter $a=1.0, 1.2, ..., 3.0$ for chain size $N=24$.
It is clear from Fig.~\ref{dimer}(a) that in the frustrated F-F model, for values of the frustration
$\alpha < \alpha_{c_{_{1}}}=\frac{1}{4}(1+a/2)$  the dimer
order parameter is equal to zero in well agreement with fully polarized ferromagnetic phase. By further increasing
the frustration and for $\alpha > \alpha_{c_{_{1}}}$, the dimer order parameter starts to increase
and reaches its saturation value ($\simeq0.5$) at $\alpha=\alpha_{c_{2}}(a)$.
At the first critical point, $\alpha=\alpha_{c_{1}}$, quantum fluctuations suppress the ferromagnetic ordering and the system
undergoes a quantum phase transition from the ferromagnetic phase into a phase with dimer ordering.
The positive value of the dimer order parameter in the region $\alpha>\alpha_{c_{1}}$, shows the dimerization
between next nearest neighbors which is named "Dimer-II". The oscillations (quasi-plateaus)
at finite $N$ in the region $\alpha_{c_{1}}<\alpha< \alpha_{c_{2}}$, are the result of level
crossing between the ground state and excited states of the model\cite{Mahdavifar08}. At the second quantum
critical point,  $\alpha=\alpha_{c_{2}}$, the ground state of the system goes into a phase with
almost fully polarized dimer state between next nearest neighbors. We have also checked the size effects on
the dimerization and numerical results are shown in the inset of Fig.~\ref{dimer}(a) with fixed
biquadratic exchange $a=2.0$  for different chain lengths $N=12, 16, 20, 24$.

In Fig.~\ref{dimer}(b), the dimer order parameter is plotted vs the frustrated parameter for a chain
size $N=24$ and different values of the biquadratic parameter $a<-2.0$. Indeed, in order to check
the nature of the classical suggested canted ferromagnetic phase, we have plotted the dimer order parameter
as a function of the frustrated parameter for fixed values of biquadratic exchanges in this region.
As it can be seen from Fig.~\ref{dimer}(b), in the region $\alpha<0$, namely nonfrustrated AF-F model, the ground state of
the system has the long-range dimerization between nearest neighbors, so called the Dimer-I phase.
In the case of the frustrated AF-AF model, as soon as the frustration increases from $\alpha_{c}$,
the dimerization order parameter starts to increase and becomes zero at almost Majumdar-Ghosh point $\alpha=0.5~\mid1+a/2\mid$.
The value of the the critical frustration, $\alpha_{c}$, depends on the biquadratic
exchange.  By more increasing the frustration from MG point, the dimerization increases very rapidly and reaches to the saturation
value ($d \simeq 0.4$). Thus, in the region of the biquadratic exchange $a<-2$, for negative values of the
frustration, the ground state is in the Dimer-I phase and by increasing the frustration, a quantum phase
transition happens at the critical positive frustration $\alpha_{c}$, from the Dimer-I phase into a phase
with dimer ordering between NNN which is named Dimer-II phase. In the inset of Fig.~\ref{dimer}(b) the
dimerization order parameter is plotted as a function of the frustration with fixed biquadratic exchange
$a=-3.0$  for different chain lengths $N=12, 16, 20, 24$. By comparing results of the different sizes it
can be conclude that there are two different dimer phase with true long-range ordering.

In the presence of biquadratic parameter $a$, at classical level spins order as spiral
structure in some part of phase diagram. It might be expectable that a part of the
broken symmetries in classical spiral spin configuration may remain to be
spontaneously broken even in the quantum regime.  The spirality or chirality in
quantum literature can be measured with vector chiral order parameter,

\begin{eqnarray}
\chi^{\gamma}&=&\frac{1}{N} \sum_{j} \left<GS\mid ({\bf S}_{j}\times {\bf S}_{j+1})^{\gamma}\mid GS\right>.
\end{eqnarray}
The vector chiral order correspondence to the spontaneous breaking
of the discrete  $Z_{2}$ symmetry about center. One should note that there are two
different quantum types of the  chiral ordered phases, gapped and gapless\cite{bb18,bb19}.
The vector chiral phase is characterized by long-range order of the vector chiral correlation defined as

\begin{eqnarray}
C^{\gamma}=\sum_{l=1}^{N}\left<GS\mid \chi_{j}~\chi_{j+l}\mid GS\right>.
\label{e6}
\end{eqnarray}
 \begin{figure}[t]
\includegraphics[width=1.0\columnwidth]{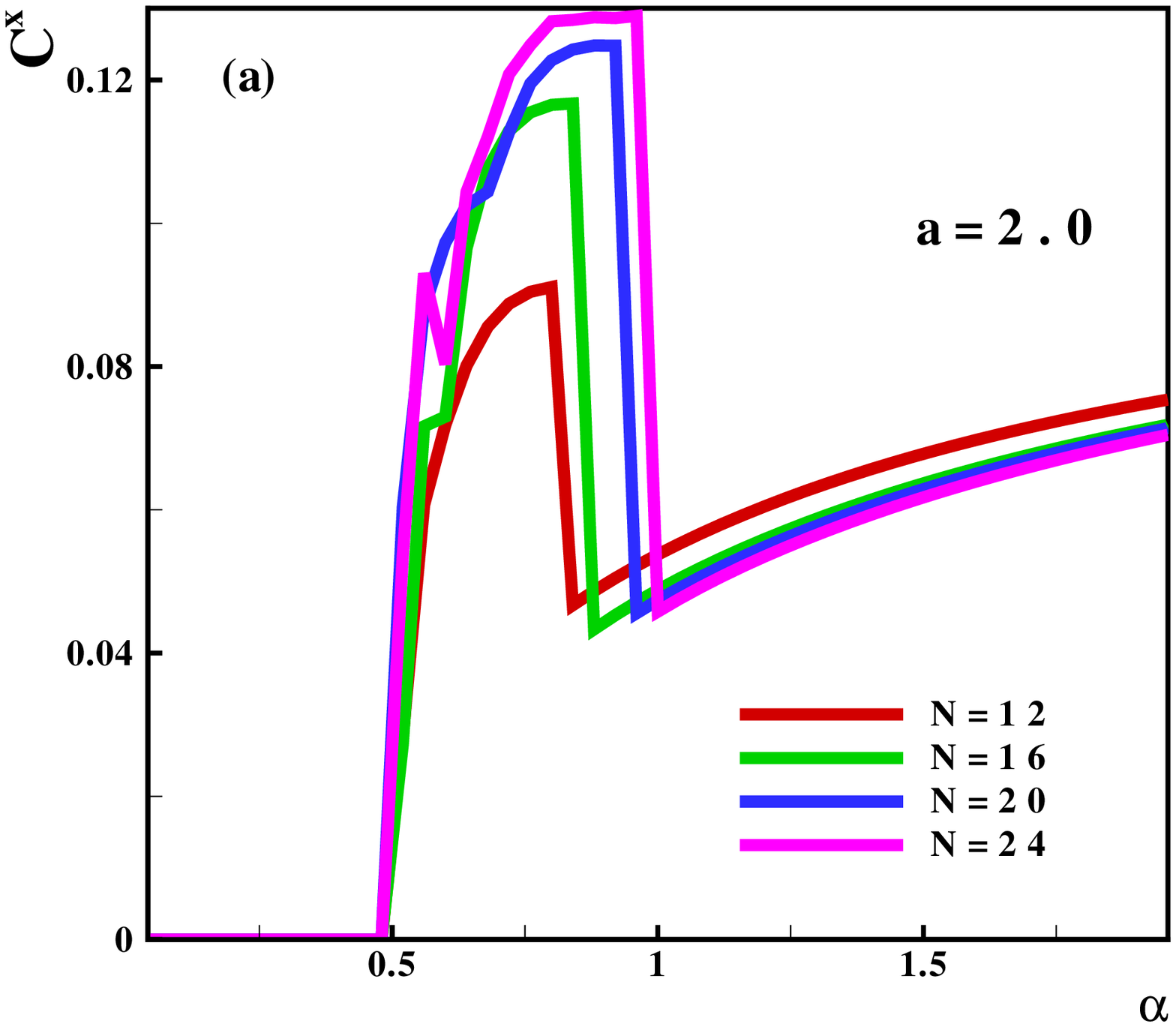}
\includegraphics[width=1.0\columnwidth]{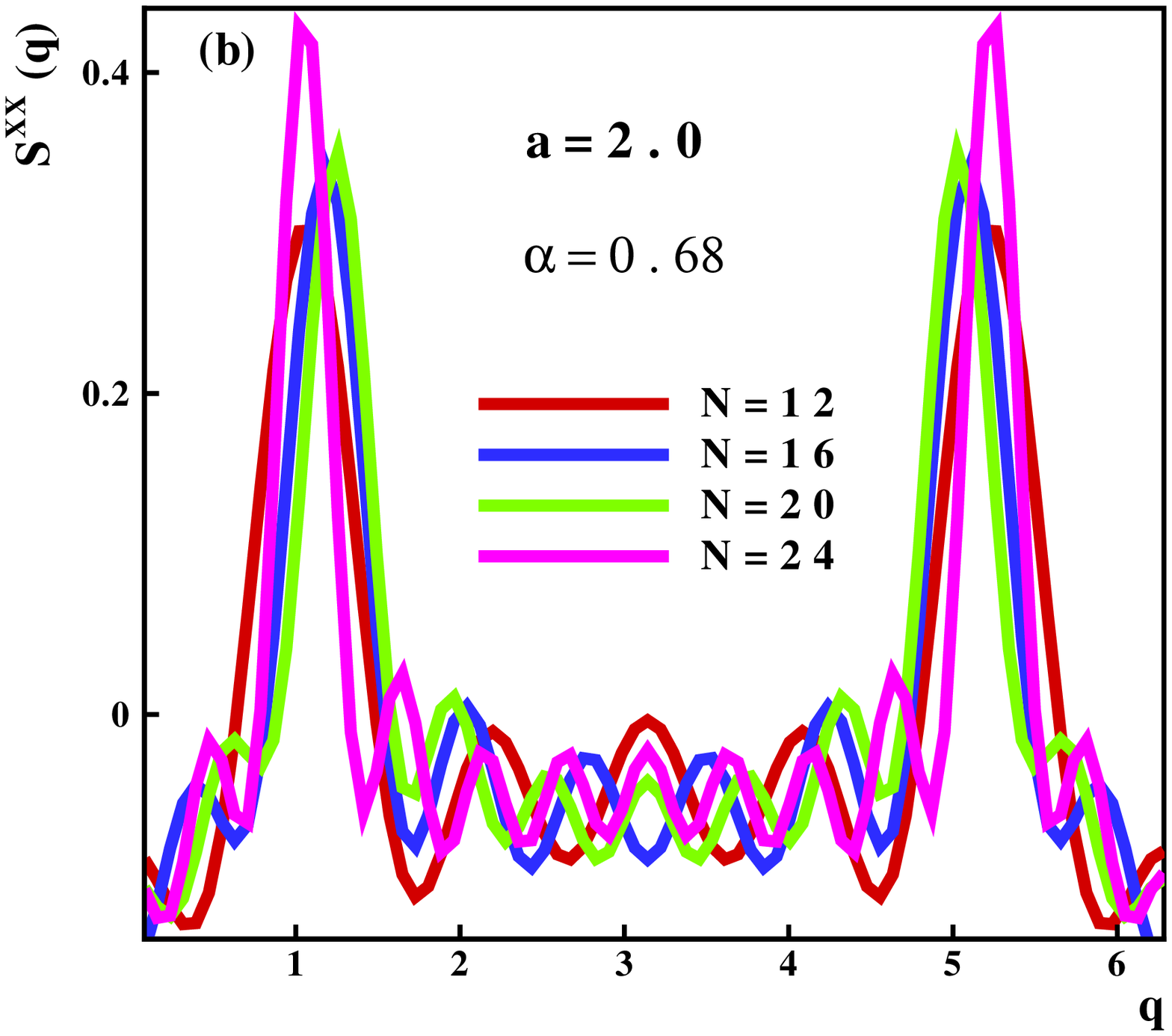}
\caption{(Color online.) (a)The vector chiral correlation as
function of frustration parameter $\alpha$
with fixed biquadratic parameter $a = 2.0$, (b) the spin structure factor as wave vector for chains with
different lengths N =12, 16, 20, 24.}
\label{chiral}
\end{figure}


To find a deeper insight into the nature of the quantum phases  we have calculated numerically the
vector chiral correlation for chains with periodic boundary conditions and lengths $N=12, 16, 20, 24$.
In Fig.~\ref{chiral}, we have presented Lanczos results on the vector chiral correlation, $C^{x}$,
as a function of the frustration parameter $\alpha$ for a fixed value of the biquadratic exchange
$a=2.0$, corresponding to the frustrated F-AF model, including different chain lengths $N=12, 16, 20, 24$.
As is clearly seen, in the region $\alpha<\alpha_{c_1}=\frac{1}{4}(1+a/2)$ there is no long-range chiral
order along the $x$ axis in well agreement with the ferromagnetic phase. By increasing the frustration,
in a intermediate region, $\alpha_{c_1}<\alpha<\alpha_{c_2}$, the ground state
shows a profound chiral order. It is important to note that the growth of the
results in the intermediate region by increasing size of the system, shows the diverging in the thermodynamic
limit $N\longrightarrow\infty$ the characteristic of the true long-range order of the chirality.
As soon as the frustration increases from $\alpha_{c_{2}}$, the chirality drops rapidly.
The constant value of the vector chiral correlation in the region $\alpha>\alpha_{c_{2}}$ shows
that the $C^{x}/N$ takes zero value in the thermodynamic limit $N\longrightarrow\infty$.
Also, we did our numerical experiment for other values of the biquadratic exchange in the
region $a>-2.0$ and found the same qualitative picture. Therefore, in the intermediate
region $\alpha_{c_1}<\alpha<\alpha_{c_2}$  and for values of the biquadratic exchange, $a>-2$,
corresponding to the frustrated F-AF model, the dimer ordering between next nearest spins coexists with the chirality.

Another way of the quantum mechanical mimic of the classical pitch angle
is the possibility to study at which wave vector $q$ the static spin structure factor
\begin{eqnarray}
S^{\alpha}(q)=\sum_{j}^{N/2}e^{iqj}\left<GS\mid S_{0}^{\alpha}S_{j}^{\alpha}\mid GS\right>.
\label{e7}
\end{eqnarray}
is peaked. In Fig.~(\ref{chiral}-b), we have plotted the structure factor versus
$0\leq q\leq2\pi$ with fixed parameters $a = 2.0$ and $\alpha=0.68$.
As it can be seen, the structure factor shows two peaks around the $q\sim1.0$ and $q\sim5.0$ in the predicted
chiral phase.

\section{Ground state entanglement} \label{sec3}
In recent years interest of the quantum information community
to study in condensed matter has stimulated an exciting cross
fertilization between the two areas \cite{bb20}. It has been found
that entanglement plays a crucial role in the low-temperature
physics of many of these systems, particularly in their ground
state\cite{bb21,bb22,bb23,bb24}. The pioneering study of quantum
information in the condensed matter area was the observation
that two body entanglement in the ground state of a cooperative
system, exhibits peculiar scaling features approaching a quantum
critical point \cite{bb22}.  These seminal studies showed that at
quantum phase transitions the dramatic change in the ground state
of a many-body system is associated to a change
in the way entanglement is distributed among the elementary
constituents. We here focus on one of the most frequently used entanglement
measure: \textit{concurrence}. A knowledge of two-site reduced density matrix
enables one to calculate concurrence, a measure of entanglement between two
spin at site $i$ and $j$ \cite{bb20,bb21}.
The reduced density matrix defined as
\begin{eqnarray}
\rho_{ij}&=&\frac{1}{4}\Big( 1+\langle\sigma_{i}^{z}\rangle\sigma_{i}^{z}+\langle\sigma_{j}^{z}\rangle\sigma_{j}^{z}+\langle\sigma_{i}^{x}\sigma_{j}^{x}\rangle\sigma_{i}^{x}\sigma_{j}^{x}\nonumber\\
&+&\langle\sigma_{i}^{y}\sigma_{j}^{y}\rangle\sigma_{i}^{y}\sigma_{j}^{y}+\langle\sigma_{i}^{z}\sigma_{j}^{z}\rangle\sigma_{i}^{z}\sigma_{j}^{z}\Big)
\label{e8}
\end{eqnarray}
where $\sigma_{i}$'s is the Pauli matrix and the concurrence $C$ is given by
$ C = max\{\varepsilon_{1}-\varepsilon_{2}-\varepsilon_{3}-\varepsilon_{4},0\}$,
where $\varepsilon_{i}$'s are square roots of the eigenvalues of the operator
$\varrho_{ij}=\rho_{ij}(\sigma_{i}^{y}\otimes\sigma_{j}^{y})\rho_{ij}^{\ast}
(\sigma_{i}^{y}\otimes\sigma_{j}^{y})$ in descending order. $C=0$ implies an
unentanglement state whereas $C=1$ corresponds to maximum entanglement.

 \begin{figure}[t]
\includegraphics[width=1.0\columnwidth]{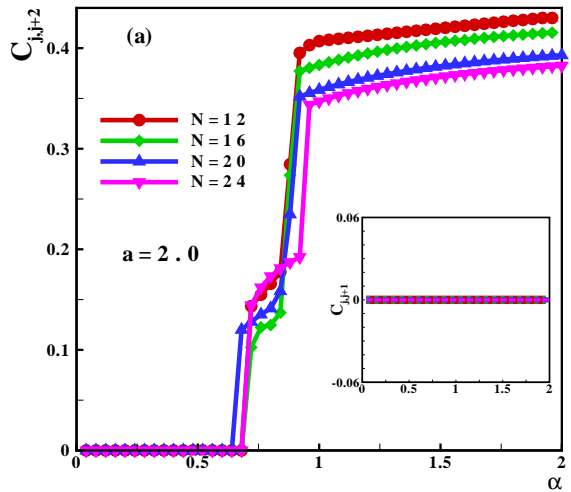}
\includegraphics[width=1.0\columnwidth]{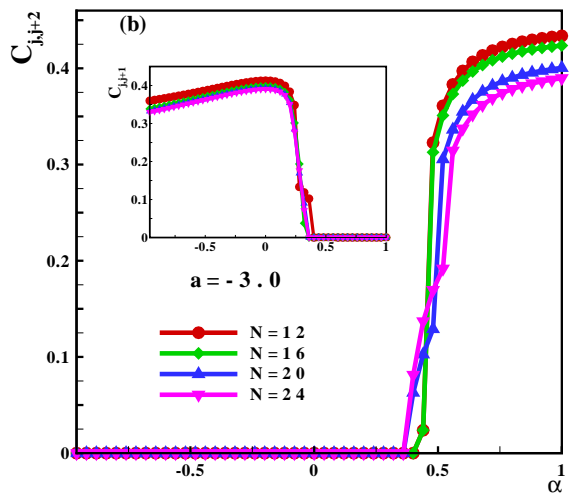}
\caption{(Color online). (a) The concurrence between next nearest neighbors $C_{j,j+2}$ as
a function of the frustration parameter $\alpha$ with a
fixed biquadratic values (a) $a=2.0$  and (b) $a=-3.0$
  for different chain lengths N =12, 16, 20, 24.
In the inset of both  plot, we plot entanglement between nearest neighbors $C_{j,j+1}$ as
a function of the frustration parameter.}
\label{Concurrence1}
\end{figure}

 \begin{figure}[t]
\includegraphics[width=1.0\columnwidth]{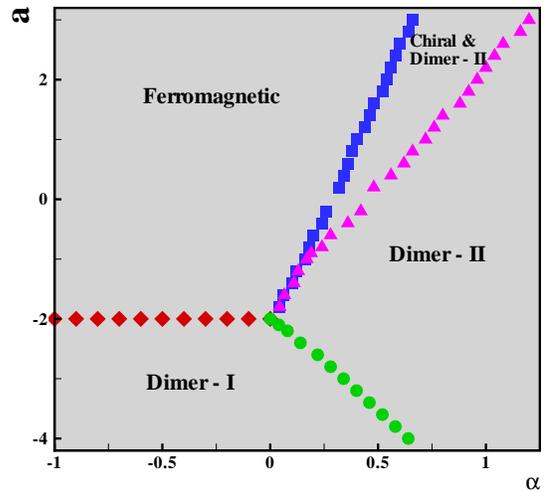}
\caption{(Color online) Modified quantum phase diagram.}
\label{schematic}
\end{figure}



The numerical Lanczos results describing the concurrence
are shown in Fig.~\ref{Concurrence1}. In this figure the
concurrence between two NN and NNN spins is plotted as a
function of the frustration $\alpha$ for chain lengths $N=12, 16, 20, 24$  with
fixed values of the biquadratic exchange. For  $a=2.0$
(Fig.~\ref{Concurrence1}(a)), corresponding to the frustrated F-AF model,
it can be seen that in the absence of the frustration, NNN spins
are not entangled in well agreement with the ferromagnetic phase.
By applying the frustration and up to the first quantum critical
point $\alpha_{c_{1}}=\frac{1}{4}(1+a/2)$, the concurrence between NNN spins remains
zero. As soon as the frustration increases from $\alpha_{c_{1}}$,
a jump happens which is the characteristic of the metamagnetic phase
transition. In the intermediate region, $\alpha_{c_1}<\alpha<\alpha_{c_2}$,
the concurrence between NNN spins increases by increasing the
frustration and reaches its nearly saturated value at $\alpha=\alpha_{c_2}$.
In the region $\alpha>\alpha_{c_2}$, the concurrence between NNN spins
remains almost constant. Indeed the quantum correlations between two NNN
spins in the intermediate region, increases with increasing the frustration
and takes the almost maximum value at $\alpha_{c_2}$. In the inset of
Fig.~\ref{Concurrence1}(a), we have plotted the concurrence between NN
spins as a function of the frustration for the biquadratic
exchange $a=2.0$. It can be seen that the NN spins do not show any entanglement in
the frustrated F-AF model.
To complete our study of the entanglement phenomena we have calculated
the concurrence between NN and NNN spins in different sectors of the
ground state phase diagram. For example, we have presented  our numerical
results for the biquadratic exchange $a=-3$ in Fig.~\ref{Concurrence1}(b).
As it can be seen, in the region of frustration, $\alpha<0$, corresponding to the nonfrustrated AF-F model,
the NNN spins are not entangled but NN spins are entangled (inset of Fig.~\ref{Concurrence1}(b)).
On the other hand, in the frustrated AF-AF model,  the NN spins remain entangled up to the Majumdar-Ghosh
point and then after the Majumdar-Ghosh by increasing the frustration parameter only
the NNN spins  will be entangled.


\section{Summary and discussion} \label{sec4}

We have considered the  frustrated ferromagnetic chains spin-$\frac{1}{2}$ with added
nearest-neighbor biquadratic interaction. In a very recent work \cite{kaplan}, the classical
ground state phase diagram of the model was studied.
The existence of ferromagnetic, spiral, canted-ferro and up-up-down-down spin structures
was shown. To find the quantum corrections, first,  using a permutation operator we eliminated
the biquadratic interaction and transformed it to the quadratic interaction. By changing the
biquadratic parameter, it is shown that the transformed Hamiltonian covers all types of NN and NNN
interaction models. Then, we did a numerical experiment to observed quantum corrections.

Our numerical experiment showed that the quantum fluctuations are strong to change the classical ground
state phase diagram. As it can be seen from Fig.~\ref{schematic}, depending on the values of the frustration
and the biquadratic exchange parameters, the ground state of the system can be found in
the ferromagnetic, the Dimer-I, the Dimer-II and the chiral magnetic orders.

In very recent works, it was shown that  the chiral phase appears in anisotropic
frustrated ferromagnetic chains and extends up to the vicinity of the SU(2) point
for moderate values of frustration\cite{saeed7, saeed9} in well agreement with the
experimental results. The complete picture of the quantum phases of this model has remained unclear up to now.
Also, several authors have discussed deeply in this area\cite{Cabra, Dmitrie, White, Allen, Nersesyan}.
The existence  of a tiny but finite gap in the region of the frustration $\alpha>0.24$ is
one of the interesting and still puzzling effects in frustrated ferromagnetic chains.

It is also worth mentioning, using the coupled cluster method for infinite chain and exact diagonalization
for finite chain, author in ref.[52], have studied the effect of a third-neighbor exchange
$J_{3}$ on the ground state of the spin half Heisenberg chain with ferromagnetic nearest-neighbor interaction $J_{1}$ and
frustrating antiferromagnetic next-nearest-neighbor interaction $J_{1}$.  By setting $J_{1}=-1$, they have proposed
that the quantum phase diagram consist of spiral and ferromagnetic phases in the $J_{2}-J_{3}$ plane. Across the $J_{3}=0$ line,
in the proposed diagram the second-order transition will take place from FM to spiral phase. Our study shows,
there should be the mentioned region and It is surprising that this region have
the two ordering phases: Dimer-II and chiral. However more research is needed in this respect.

From quantum entanglement point of view, difference between quantum phases is also studied.
we have calculated the concurrence between two NN and NNN spins in different sectors of the
ground state phase diagram. We showed that the concurrence function is a very useful tool to
recognize the different quantum phases specially in this model.

\section{Acknowledgement}
It is our pleasure to thank T. Vekua and T. Nishino, for very useful comments.

\vspace{0.3cm}

\end{document}